\documentclass{jnmp}
\usepackage{graphicx,dcolumn,amsmath,amssymb,amsfonts,epsfig,float,subfigure}
\usepackage{amsthm}
\usepackage{txfonts}
\usepackage[T1]{fontenc}
\usepackage[english]{babel}
\usepackage{bm}

\JNMPnumberwithin{equation}{section}
\resetfootnoterule
\theoremstyle{definition}

\newtheorem{acknowledgments}{Aknowledgments}[section]

\begin{document}
\renewcommand{\evenhead}{A Bourlioux, R Rebelo and P Winternitz}
\renewcommand{\oddhead}{Symmetry preserving discretization of SL(2,$\mathbb{R}$) invariant equations}
\thispagestyle{empty}
\FirstPageHead{*}{*}{20**}{\pageref{firstpage}--
\pageref{lastpage}}{Article}
\copyrightnote{200*}{A Bourlioux, R Rebelo and P Winternitz}
\Name{Symmetry preserving discretization of SL(2,$\mathbb{R}$) invariant equations}
\label{firstpage}
\Author{Anne Bourlioux~$^a$, Raphaël Rebelo~$^{b,c}$ and Pavel Winternitz~$^{a,b}$}
\Address{$^a$ Département de mathématiques et de statistique, Université de Montréal, C.P. 6128, succ. Centre-ville, Montréal, Québec H3C 3J7, Canada.\\
~~E-mail: bourliou@dms.umontreal.ca\\[10pt]
$^b$ Centre de recherches mathématiques, Université de Montréal, C.P. 6128, succ. Centre-ville, Montréal, Québec H3C 3J7, Canada.\\
~~E-mail: wintern@crm.umontreal.ca\\[10pt]
$^c$ Département de Physique, Université de Montréal, C.P. 6128, succ. Centre-ville, Montréal, Québec H3C 3J7, Canada.\\
~~E-mail: raph.rebelo@gmail.com\\[10pt]}
\Date{Received Month *, 200*; Accepted Month *, 200*}
\begin{abstract}
\noindent
Nonlinear ODEs invariant under the group SL(2,$\mathbb{R}$) are solved numerically. We show that solution methods incorporating the Lie point symmetries provide better results than standard methods.
\end{abstract}

\newpage
\section{Introduction}

Historically Lie group theory started out as a theory of transformations of solutions of ordinary and partial differential equations. Differential equations are still one of the most important applications of Lie groups \cite{1}. The most common use of Lie group theory in this field is to perform symmetry reduction. For ordinary differential equations (ODE), this means that Lie point symmetries are used to reduce the order of the equation. If the symmetry group is large enough the order of the ODE can be reduced to zero. This is equivalent to obtaining the general solution of the ODE, possibly in implicit form. For partial differential equations (PDE), symmetry reduction means a reduction of the number of independent variables and usually leads to exact analytical solutions, albeit particular ones.\\

The purpose of this article is to discuss a different application of Lie groups in the theory of differential equations and to present some new results and new examples. This application can be called {\it symmetry preserving discretization of differential equations} and its purpose is to improve numerical methods for solving differential equations. In this article we restrict ourselves to ODEs. For recent reviews with references to original articles see \cite{2,3}.\\

The idea is to start from a given ODE of order N
\begin{gather}\label{ODEintro}
F(x,y,y',y'',...,y^{(N)})=0
\end{gather}
and its known Lie point symmetry group G with a Lie algebra L (the symmetry algebra) realized by vector fields of the form
\begin{gather}\label{vectorintro}
X=\xi(x,y)\partial_x+\phi(x,y)\partial_y
\end{gather}
\eqref{ODEintro} is replaced by an invariant difference scheme, i.e. a system of two equations
\begin{align}\label{schemeintro}
E_a&(n,x_{n+K},x_{n+K+1},...,x_{n+L},y_{n+K},y_{n+K+1},...,y_{n+L})=0\\
a&=1,2 \qquad L-K=N, \qquad N \geq M\nonumber
\end{align}
relating the variables x and y in M different points.\\

The scheme \eqref{schemeintro} is constructed so as to be invariant under the same group G as the ODE \eqref{ODEintro}. This means that equations \eqref{schemeintro} must be annihilated on their solution set by the prolongations of the vector fields \eqref{vectorintro}
\begin{gather}
pr X(E_a)|_{E_1=E_2=0}=0
\end{gather}
where the {\it discrete prolongation} is 
\begin{gather}
prX=\sum_{i=n+K}^{n+L}{\left\{\xi(x_i,y_i)\partial_{x_i}+\phi(x_i,y_i)\partial_{y_i}\right\}}
\end{gather}

In practise, this means that we can proceed as follows:\\

1. Find the N-th order differential invariants $I_j^c(x,y,y',...,y^(N))$, $j=1,...,J$, of the group G and rewrite the ODE \eqref{ODEintro} in terms of these invariants :
\begin{gather}
\text{\~{F}}(I_1^c,I_2^c,...,I_J^c)=0
\end{gather}

2. Find the difference invariants $I_a(n,x_{n+K},...,x_{n+L},y_{n+K},...,y_{n+L})$ of the same group G.\\

3. Expand the difference invariants in Taylor series about some reference point, say $(x_n,y_n)$. Choose such difference invariants that we have
\begin{gather}
I_j=I_j^c+0(\epsilon)
\end{gather}
i.e. such that the leading terms in the expansion coincides with a corresponding differential invariant ($\epsilon$ $\to$ 0 is the continuous limit).\\

4. Write the difference scheme \eqref{schemeintro} in terms of difference invariants $I_j$. In the continuous limit we will have
\begin{align}
\text{\~{E}}_1(I_1^c,...,I_J^c)&=0 \quad \stackrel{\epsilon \to 0}{\longrightarrow} \quad \text{\~{F}}(I_1^c,I_2^c,...,I_J^c)=0\\
\text{\~{E}}_2(I_1^c,...,I_J^c)&=0 \quad \stackrel{\epsilon \to 0}{\longrightarrow} \quad 0=0\nonumber
\end{align}

In the invariant discretization the lattice is not given a priori, but emerges as part of the solution of the difference scheme.\\

In physics and other fields of science, symmetries of a system are often better known than the dynamics and symmetries of equations can be more important than explicit solutions. Preserving symmetries in numerical calculations can be expected to improve the results, specially global features of solutions.\\

The invariant discretization should be compared to what we will call {\it standard discretizations}. The lattice is given a priori, usually a uniform one. The ODE \eqref{ODEintro} is discretized by replacing all derivatives by usual finite differences, e.g.,
\begin{gather}
u_x=\frac{u(x_{n+1})-u(x_n)}{x_{n+1}-x_n}, \quad x_n=nh+x_0
\end{gather}
and similarly for higher order derivatives.\\

Recent articles devoted to theoretical aspects of invariant discretization of ODEs include \cite{4,5,6}. In Ref.\cite{7} it was shown on several examples that the symmetry preserving schemes provide better accuracy than standard ones and numerical solutions close to singularities where standard schemes fail. Here we will show that similar results hold in other situations.

\section{A two dimensional realization of the algebra sl(2,$\mathbb{R}$)}

\subsection{The Lie algebra and the invariant ODEs}

Let us consider the sl(2,$\mathbb{R}$) algebra with a basis realized by the vector fields
\begin{gather}\label{algebra}
X_1=\partial_y, \qquad X_2=x\partial_x+y\partial_y, \qquad X_3=2xy\partial_x+y^2\partial_y \quad.
\end{gather}
It can be extended to a gl(2,$\mathbb{R}$) algebra by adding
\begin{gather}
X_4=y\partial_y
\end{gather}
to the basis.  Let us prolong these vector fields so they act on functions $F(x,y,y',y'',y''')$.  The action of the corresponding SL(2,$\mathbb{R}$) group on the prolonged space with local coordinates $\{x,y,y',y'',y''' \}$ allows two differential invariants, namely
\begin{gather}\label{continuousinvariants}
I_1^c=\frac{2xy''+y'}{y'^3}, \qquad I_2^c=\frac{x^2(y'y'''-3y''^2)}{y'^5} \quad.
\end{gather}

Using these invariants we can write a second and a third order invariant ODE, namely
\begin{gather}\label{ODE2ndorder}
2xy''+y'=\gamma y'^3
\end{gather}
and
\begin{gather}\label{ODE3rdorder}
\frac{x^2(y'y'''-3y''^2)}{y'^5}=F(I_1^c)
\end{gather}
where $\gamma$ is a constant and F(z) is an arbitrary function.  The ODE \eqref{ODE3rdorder} will be invariant under the group GL(2,$\mathbb{R}$), including the dilatations generated by $X_4$ if we restrict $F(z)$ to be $F(z)=\alpha z^{3/2}$.  Eq. \eqref{ODE3rdorder} specializes to
\begin{gather}\label{ODE3rdordergl}
x^2(y'y'''-3y''^2)=\alpha(2xy''+y')^{3/2}y'^{1/2}
\end{gather}
where $\alpha$ is a constant.\\

By construction the ODEs \eqref{ODE2ndorder} and \eqref{ODE3rdordergl} have symmetry algebras that make it possible to reduce them to quadratures.\\

For eq. \eqref{ODE2ndorder} this provides two explicit solutions
\begin{gather}\label{ODE2ndordersolution}
y_{1,2}(x)=
\left\{
\begin{aligned}
&y_b \pm \frac{2}{C}\sqrt{C-\gamma x} \qquad &C\neq 0\\
&y_b \pm \frac{1}{\sqrt{\gamma}}x \qquad &C=0 .
\end{aligned}
\right.
\end{gather}

The two branches of the solution for C $\neq$ 0 intersect for $x=C/\gamma$ where we have $y_1=y_2=y_b$. After they become complex, the solution for $x=C/\gamma$ remains finite but all its derivatives become infinite.\\

For eq. \eqref{ODE3rdordergl} the quadratures lead to an implicit solution :
\begin{gather}\label{3rdordery}
y=y_0+C_1\int_0^x{e^{\int_0^t{f(s)ds}}dt}
\end{gather}
where f(x) satisfies
\begin{gather}\label{3rdorderfunction}
f(x)=\frac{1}{2x}\left[ \frac{1}{Kx}\left( \frac{\sqrt{2xf(x)+1}+\alpha-\sqrt{\alpha^2+1}}{\sqrt{2xf(x)+1}+\alpha+\sqrt{\alpha^2+1}} \right)^{\left(\frac{\alpha-\sqrt{\alpha^2+1}}{\sqrt{\alpha^2+1}} \right)}-1 \right]
\end{gather}
(K$\neq$0, $C_1$ and $y_0$ are constants).\\

Eq. \eqref{ODE3rdordergl} provides a good example.  The symmetry group is large enough to reduce to quadratures. This however really means that we have replaced a differential equation for y(x) by a functional equation \eqref{3rdorderfunction} for f(x).  To obtain a graph $y=y(x)$ we still have to do numerical calculations.

\subsection{The difference invariants}

Let us consider four points $x_k$ on a line and the values $y_k=y(x_k)$ at these points :
\begin{gather}\label{lattice}
(x_{n-1},x_n,x_{n+1},x_{n+2},y_{n-1},y_n,y_{n+1},y_{n+2}) \quad.
\end{gather}

The SL(2,$\mathbb{R}$) group generated by the prolongation of the vector fields \eqref{algebra} to the points \eqref{lattice} will transform these points in the (x,y) plane but will leave certain functions of them invariant.  We calculate these invariants using known methods [2,...,7]. The result is that out of these coordinates we can construct five 4-point difference invariants
\begin{gather}
I_1^n=\frac{y_n-y_{n-1}}{\sqrt{x_nx_{n-1}}}, \qquad I_1^{n+1}=\frac{y_{n+1}-y_{n}}{\sqrt{x_{n+1}x_{n}}}, \qquad I_2^{n+1}=\frac{y_{n+1}-y_{n-1}}{\sqrt{x_{n+1}x_{n-1}}}\\
I_1^{n+2}=\frac{y_{n+2}-y_{n+1}}{\sqrt{x_{n+2}x_{n+1}}}, \qquad I_2^{n+2}=\frac{y_{n+2}-y_{n}}{\sqrt{x_{n+2}x_{n}}}\quad.
\end{gather}

We mention that $I_1^{n+1}$, $I_1^{n+2}$ are just upshifts of $I_1^n$, $I_2^{n+2}$ is an upshift of $I_2^{n+1}$.  Moreover, $I_1^n$, $I_1^{n+1}$ and $I_2^{n+1}$ involve coordinates of the first three points only.\\

Let us now obtain difference schemes for the ODEs \eqref{ODE2ndorder} and \eqref{ODE3rdordergl}.  We put
\begin{gather}
h_n=x_n-x_{n-1}, \qquad h_{n+1}=x_{n+1}-x_n, \qquad h_{n+2}=x_{n+2}-x_{n+1}
\end{gather}
and expand $y_{n+k}\equiv y(x_{n+k})$ about some point $x_0$.  First of all, we notice that the equation
\begin{gather}\label{2ndorderlattice}
I_1^{n+1}-I_1^n=0
\end{gather}
provides a good lattice.  Indeed expanding \eqref{2ndorderlattice} about the point $x_n$ we obtain
\begin{gather}\label{2ndorderlatticeexpansion}
I_1^{n+1}-I_1^{n}=\frac{y'}{x}(h_{n+1}-h_n)+\frac{xy''-y'}{x^2}(h_{n+1}^2+h_n^2)+...
\end{gather}

In the continuous limit we put
\begin{gather}
h_n=\alpha_n\epsilon
\end{gather}
where $\alpha_n$ are constants of the order $\alpha_n\sim1$ and take $\epsilon \rightarrow 0$.  From eq. \eqref{2ndorderlattice} and \eqref{2ndorderlatticeexpansion}, we see that for $\epsilon \rightarrow 0$ we have
\begin{gather}\label{latticeorder}
\alpha_{n+1}-\alpha_{n}\sim 0(\epsilon)
\end{gather}
(if $\frac{xy''-y'}{2y'x}$ is finite)\\

Let us now approximate the continuous invariants $I_1^c$ and $I_2^c$ of eq. \eqref{continuousinvariants}.  To obtain $I_1^c$, we need 3 points $( n-1,n,n+1)$ or $( n,n+1,n+2)$.  From the expansion of $I_1^{n-k}$ and $I_1^{n+k}$, we see that the correct expression is
\begin{align}\label{2ndorderschemeexpansion}
J_1^{n+1}&=8\frac{I_2^{n+1}-(I_{1}^n+I_1^{n+1})}{I_1^nI_{1}^{n+1}(I_{1}^{n}+I_1^{n+1})}
   &=\frac{2xy''+y'}{y'^3}+2(h_{n+1}-h_n)x\frac{-3y''^2+y'y'''}{3y'^4}+0(h^2)
\end{align}
and similarly for $J_1^{n+2}$ (an upshift of $J_1^{n+1}$).  We can now consider $\{ I_1^n,I_1^{n+1},I_1^{n+2},J_1^{n+1},J_1^{n+2} \}$ as a basis for the four point difference invariants.  Both $J_1^{n+1}$ and $J_1^{n+2}$ have the differential invariant $I_1^c$ as their continuous limit.  Moreover, on the lattice \eqref{2ndorderlattice}, this is an approximation of the order $\epsilon^2$ (see \eqref{2ndorderlatticeexpansion}, \eqref{latticeorder}, \eqref{2ndorderschemeexpansion}).\\

To approximate $I_2^c$ we must use all 4 points.  Indeed, we have
\begin{gather}\label{3rdorderschemeexpansion}
K^{n+2}=\frac{3}{2}\left(\frac{J_1^{n+2}-J_1^{n+1}}{I_1^n+I_1^{n+1}+I_1^{n+2}}\right)=\frac{x^2}{y'^5}(y'y'''-3y''^2)+(h_{n+2}-h_n)R(x,y,y',y'',y''',y^{(4)})
\end{gather}
where R is some differential expression (that is easy to calculate).  Thus $K^{n+2}$ goes to $I_2^c$ for $\epsilon \rightarrow 0$ and provides an approximation of order $\epsilon^2$ on the lattice \eqref{2ndorderlattice} (since \eqref{2ndorderlattice} implies also $I_1^{n+2}=I_1^{n+1}=I_1^{n}\equiv I_1$ with $I_1$ independent of n).

\subsection{Difference schemes for the second order equation}

We put
\begin{gather}\label{2ndorderscheme}
J_1^{n+1}=\gamma, \qquad I_1^{n+1}=I_1^n\equiv I \quad.
\end{gather}

From eq. \eqref{2ndorderscheme} we obtain
\begin{gather}
I_2^{n+1}=I_1(\frac{\gamma}{4}I_1^2+2)\equiv \beta \quad.
\end{gather}

Since $I_1$ is a constant (depending on the initial conditions $(x_0,y_0,x_1,y_1)$) $\beta$ will also be a constant.\\

The scheme \eqref{2ndorderscheme} can be solved explicitly for $x_{n+1}$ and $y_{n+1}$ and we obtain
\begin{gather}\label{2ndorderschemeexplicit}
x_{n+1}=x_{n-1}\left[\frac{y_n-y_{n-1}}{\beta x_{n-1}-(y_n-y_{n-1})}\right]^2 \quad , \quad  y_{n+1}=\frac{\beta x_{n-1}y_n-(y_n-y_{n-1})y_{n-1}}{\beta x_{n-1}-(y_n-y_{n-1})} \quad.
\end{gather}

Thus, the invariant scheme is an explicit and linear one and moreover it is of the order $\epsilon^2$.  By comparison, a standard scheme of order $\epsilon^2$ will be implicit and $y_{n+1}$ will be obtained by solving a cubic equation.  An explicit standard scheme will be of order $\epsilon$. We mention here that it is the difference scheme that converges to the ODE like $\epsilon^2$. This does not guarantee that the same is true for the solutions.

\subsection{Difference scheme for the third order ODE}

From eq. \eqref{3rdorderschemeexpansion} we see that an invariant difference scheme for the third order equation \eqref{ODE3rdorder} is obtained by putting
\begin{gather}\label{3rdorderscheme}
K^{n+2}=F(J_1^{n+1}), \qquad I_1^{n+2}=I_1^{n+1}=I_1^n\equiv I \quad.
\end{gather}

Alternatively, we can put
\begin{gather}\label{3rdorderschemeimplicit}
K^{n+2}=F\left(\frac{J_1^{n+1}+J_1^{n+2}}{2}\right), \qquad I_1^{n+2}=I_1^n\equiv I \quad.
\end{gather}

Both schemes converge to the ODE like $\epsilon^2$.  The scheme \eqref{3rdorderscheme} can be solved explicitly for $x_{n+2}$, $y_{n+2}$ and we obtain
\begin{gather}\label{3rdorderschemeexplicit}
x_{n+2}=\frac{x_n}{(1-\omega_{n+1})^2} \qquad y_{n+2}=\frac{y_n-\omega_{n+1}y_{n+1}}{1-\omega_n}
\end{gather}
where we have defined
\begin{gather}
\omega_{n+1}=\frac{I_1^2}{4}[2I_1F(J_1^{n+1})+J_1^{n+1}]\sqrt{\frac{x_n}{x_{n+1}}} \quad.
\end{gather}

We stress that $I_1$ is a constant (independent of n and determined by the initial conditions) but $J_1^{n+1}$ and hence $\omega_{n+1}$ depend on n and must be calculated at each step using
\begin{gather}
J_1^{n+1}=\frac{4}{\sqrt{x_{n+1}x_{n-1}}I_1^3}[y_{n+1}-y_{n-1}-2I_1\sqrt{x_{n+1}x_{n-1}}]
\end{gather}

Thus the SL(2,$\mathbb{R}$) invariant scheme is a very simple one for any function $F(J_1)$:  $x_{n+2}$ and $y_{n+2}$ are obtained explicitly in terms of their values at $n-1$, n and $n+1$.  The standard scheme will be nonlinear and hence implicit. The condition $I_1^n=I_1^{n+1}$ in \eqref{3rdorderscheme} is actually quite restrictive for third (and higher) order ODEs. The constant I is determined by the initial conditions and \eqref{3rdorderscheme} imposes a relation between $y(0)$, $y'(0)$ and $y''(0)$. For general initial conditions a relation of the type $I_1^{n+1}=\gamma I_1^{n}$ is more suitable ($\gamma$ is determined in terms of the initial conditions).

\section{One dimensional realization of sl(2,$\mathbb{R}$)}

\subsection{The Lie algebra and differential invariants}

The Lie algebra is realized by the vector fields
\begin{gather}\label{algebra1}
X_1=\partial_y, \qquad X_2=y\partial_y, \qquad X_3=y^2\partial_y
\end{gather}
so x is an invariant quatity.  The lowest order differential invariant is the Schwartzian derivative
\begin{gather}
I_1=\frac{1}{y'^2}(y'y'''-\frac{3}{2}y''^2) \quad.
\end{gather}

Thus the equation
\begin{gather}\label{ode}
\frac{1}{y'^2}(y'y'''-\frac{3}{2}y''^2)=F(x)
\end{gather}
will be invariant under this realization of SL(2,$\mathbb{R}$) for any function $F(x)$.  For $F(x)=$const. eq \eqref{ode} is invariant under GL(2,$\mathbb{R}$) generated by \eqref{algebra1} and $X_4=\partial_x$.  For $F(x)=0$ the ODE is invariant under SL(2,$\mathbb{R}$)$\otimes$SL(2,$\mathbb{R}$).

\subsection{Invariant difference scheme}

The ODE \eqref{ode} can be approximated by a four point difference scheme.  The space of difference invariants is five dimensional and is generated by
\begin{gather}
R^{n+2}=\frac{(y_{n+2}-y_n)(y_{n+1}-y_{n-1})}{(y_{n+2}-y_{n+1})(y_{n+1}-y_n)}, \quad x_{n-1}, \quad x_n, \quad x_{n+1}, \quad x_{n+2}
\end{gather}

and the ODE \eqref{ode} is approximated by
\begin{gather}\label{scheme}
J^{n+2}=\frac{4-R^{n+2}}{2h^2}=F(x_n,h)\\
h_{n+1}=h_n=h, \qquad F(x_n,0)=F(x)
\end{gather}

We have chosen a uniform lattice but any equation $E(x_{n-1},x_n,x_{n+1},x_{n+2})=0$ will provide an alternative invariant lattice.\\

The scheme \eqref{scheme} can be solved explicitly and we obtain
\begin{gather}\label{explicitscheme}
x_{n+2}=x_{n+1}+h=(n+2)h+x_0 \quad , \quad y_{n+2}=\frac{K_n(y_n-y_{n-1})y_{n+1}-(y_{n+1}-y_{n-1})y_n}{K_n(y_n-y_{n-1})-(y_{n+1}-y_{n-1})}
\end{gather}
where we have put
\begin{gather}
K_n=4\left[1-\frac{h^2}{2}F\left(x_n+\frac{h}{2}\right)\right] \quad.
\end{gather}
Expanding $J^{n+2}$ in \eqref{scheme} about the point $x_n+\frac{h}{2}$ we obtain
\begin{gather}
K_n=\frac{1}{y'^2}(y'y'''-\frac{3}{2}y''^2)+0(h^2) \quad.
\end{gather}

Thus the scheme \eqref{explicitscheme} approximates the ODE \eqref{ode} with second order accuracy and the scheme is explicit and linear.\\

A standard scheme which is second order accurate will be implicit and $y_{n+2}$ will be calculated from an algebraic equation (at least quadratic).

\section{Numerical analysis}

In this section we concentrate on the two-dimensional realization of sl(2,$\mathbb{R}$) and apply the symmetry preserving and standard schemes to \eqref{ODE2ndorder} and \eqref{ODE3rdorder}. For some numerical results concerning the one-dimensional realization of sl(2,$\mathbb{R}$) see Ref.\cite{7}.

\subsection{Second order equations}

The general solution of \eqref{ODE2ndorder} is given in \eqref{ODE2ndordersolution} for $C \neq 0$. On Fig.\ref{fig1} we show the exact solutions $y_1$ (increasing branch) and $y_2$ (decreasing branch) for $\gamma=150$, $y_b=5$, $C=e^2$. The step for the exact solution was $h=0.05$. The symmetry preserving method integrates $y_1$ up to the singularity at $x=x_0=C/\gamma \sim 20.3$ then continues along the second branch $y_2$ to its 'initial' value. The standard method fails to converge close to the singularity (where the solution becomes complex).

\subsection{Third order equation}

Let us now consider eq. \eqref{ODE3rdordergl} with $\alpha=-1$. We put $I_1^{n+1}=\gamma I_1^n$ where $\gamma$ is determined by the initial conditions. As the step h tends to zero, $\gamma$ will tend to $\gamma=1$ as required in eq. \eqref{2ndorderlattice}.\\

On Fig.\ref{fig2} we compare the accuracies of the standard and symmetry preserving scheme. Since in this case no exact analytic solution is available we compare with a reference solution using a Matlab Runge-Kutta scheme with a tolerance on the error set at tol$=10^{-9}$. The initial conditions were set at $y_0=1$, $y_0'=10$, $y_0''=-4$ and this corresponds to a solution with no singularity on the real axis for $0\leq x \leq 16$. We see that the accuracy is better for the symmetry preserving scheme by a factor of 10.\\

A singular solution is shown on Fig.\ref{fig3}. The initial conditions were set at $y_0=1$, $y_0'=1$, $y_0''=3$ and a singularity occurs for $x \sim 1.7$. Matlab solvers and standard schemes stop providing solutions close to the singularity. The symmetry preserving method approaches the singularity closely and then continues along the second branch of the singular solution towards an appropriate initial condition. Qualitatively we have the same features as for the second order equation \eqref{ODE2ndorder}. The solution itself stays finite but its derivative becomes infinite at the singularity.

\section{Conclusions}

The main conclusion that we draw from the examples so far considered is that the symmetry preserving schemes are actually simpler than the standard ones. We have shown that they can be explicitly solved (see eq. \eqref{2ndorderschemeexplicit} and \eqref{3rdorderschemeexplicit}) while preserving second order accuracy. Not surprisingly, the numerical calculations confirm that the symmetry preserving schemes have better accuracy and provide better results close to singularities.\\

\begin{figure}[htbp]
\begin{center}
\includegraphics[width=12cm,height=9cm]{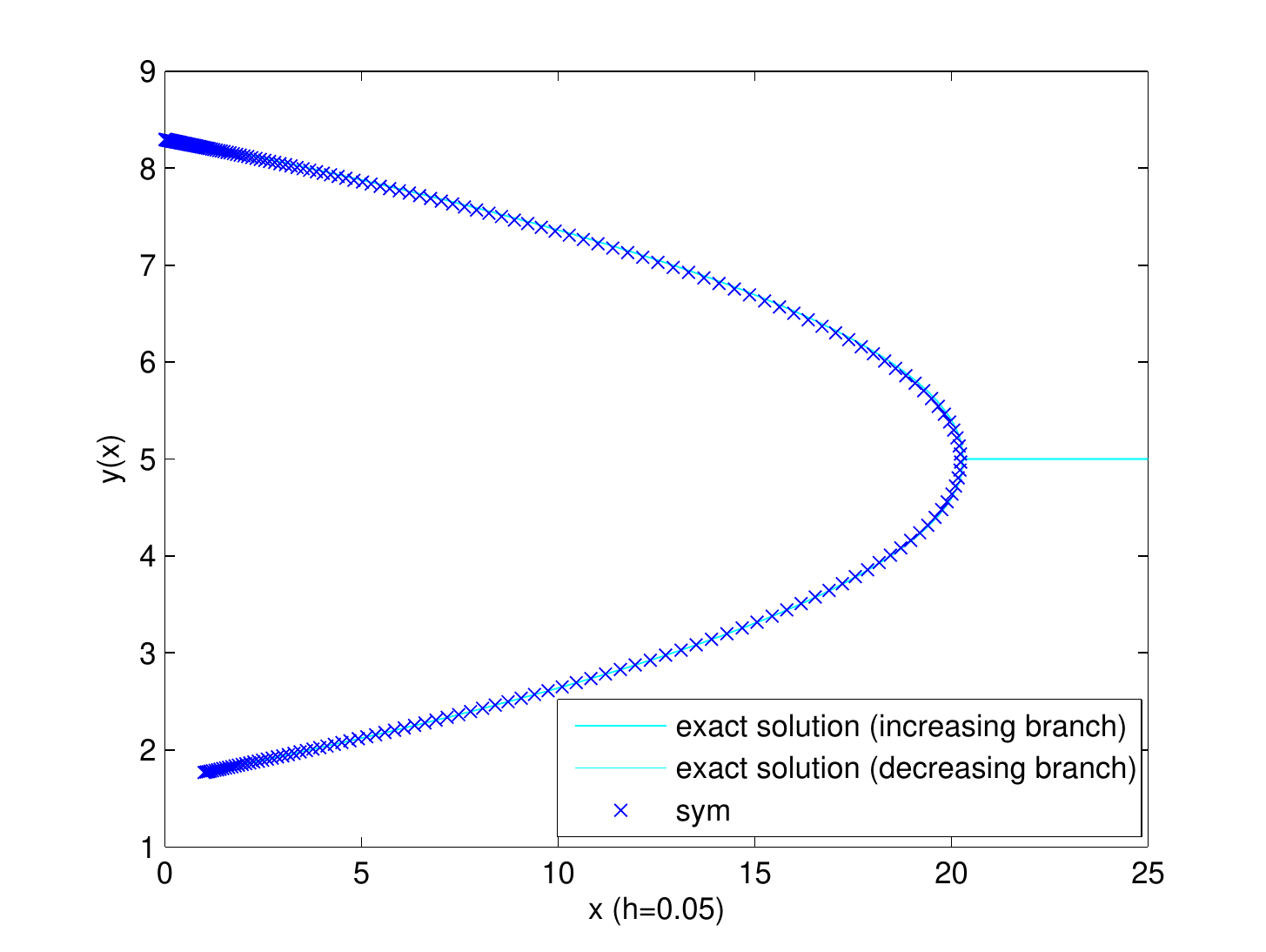}
\caption{Behaviour of the symmetry preserving scheme near the singularity for eq. \eqref{ODE2ndorder}}
\label{fig1}
\end{center}
\end{figure}

\begin{figure}[htbp]
\begin{center}
\includegraphics[width=15cm,height=20cm]{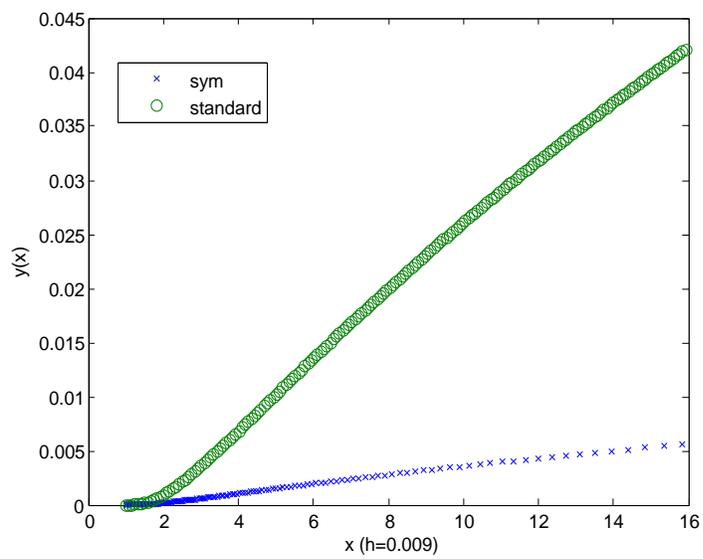}
\caption{Discretization errors for standard and symmetry preserving schemes for eq. \eqref{ODE3rdordergl}, $\alpha=-1$ for a regular solution}
\label{fig2}
\end{center}
\end{figure}

\begin{figure}[htbp]
\begin{center}
\includegraphics[width=15cm,height=20cm]{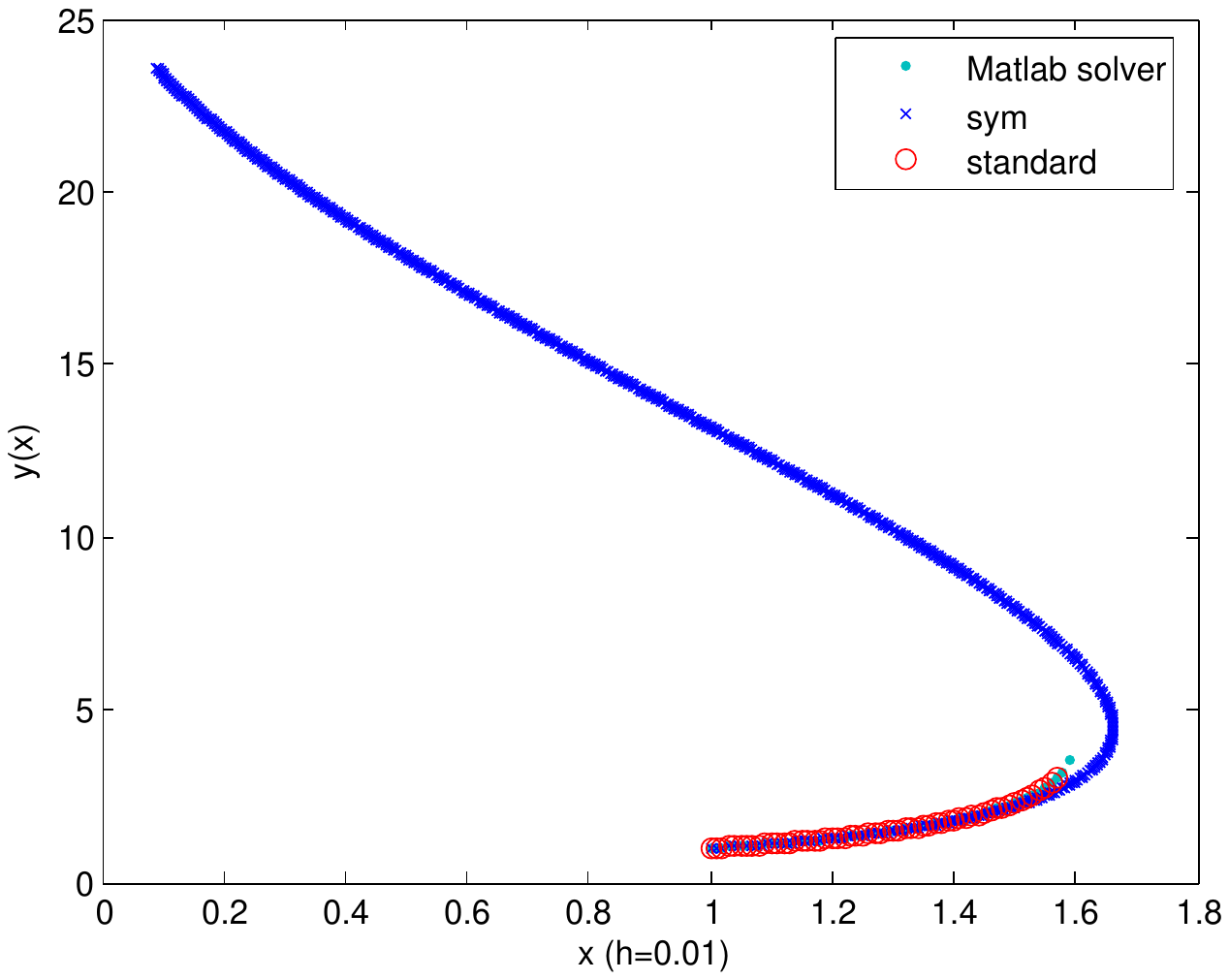}
\caption{Behaviour of the symmetry preserving scheme near a singularity for eq. \eqref{ODE3rdordergl}}
\label{fig3}
\end{center}
\end{figure}

A specific conclusion from Fig.\ref{fig1} and Fig.\ref{fig3}, born out by other examples \cite{7}, is that the symmetry preserving method provides qualitative information about singularities of the solutions, not available from standard methods. We can not only pinpoint the position of the singularity. We also see from the curves that the solution itself remains finite, but its derivative becomes infinite. In addition we see that it is a square root type singularity. Indeed if we follow the solution backwards from the singularity at $x=x_b$ we see two different branches. For the second order equation \eqref{ODE2ndorder} this behaviour is obvious from the explicit solution \eqref{ODE2ndordersolution}. For the third order equation \eqref{ODE3rdordergl} this is not visible without numerical calculations.\\

\begin{acknowledgments}
The research of A.B. and P.W. was supported by research grants from NSERC of Canada. P.W. and R.R. thank the Universita Roma-Tre for hospitality and D.Levi for helpful discussion.
\end{acknowledgments}

\label{lastpage}

\end{document}